\documentstyle[aps,prl,epsf,floats]{revtex} 
\begin{document} 
\twocolumn[\hsize\textwidth\columnwidth\hsize\csname@twocolumnfalse\endcsname 
\title{Kondo effect in ``bad metals''} 
\author{S. Burdin$^{1,2}$, V. Zlati\'c$^{3}$} 
\address{ 
$^{1}$ Institut Laue-Langevin, B.P. 156, 38042 Grenoble Cedex 9, France} 
\address{
$^2$ INFM e Dipartimento di Fisica, 
Universit\`a di Roma, ``La Sapienza'', Piazzale A. Moro 2, 00185 Roma, Italy
}
\address{
$^{3}$ Institute of Physics, P.O. Box 304, 10001 Zagreb, Croatia} 
\date{\today} 
\maketitle 
\widetext 
\begin{abstract}
We study the low-temperature properties of a Kondo lattice using the 
large$-{\cal N}$ formalism. For a singular density of conduction states 
(DOS), we generalize the single-impurity result of Withoff and Fradkin: the 
strong-coupling fixed point becomes irrelevant if the DOS vanishes at the 
Fermi level $E_F$. However, for $E_F$ close enough to the singularity, and 
close to half-filling, the Kondo temperature, $T_K$, can become  much smaller 
than the characteristic Fermi liquid scale. At T=0, a meta-magnetic 
transition occurs at the critical magnetic field $H_c\simeq (k_B/\mu_B) T_K$. 
Our results provide a qualitative explanation for the behavior of the 
YbInCu$_4$ compound below the valence-change transition. 
\end{abstract}
] 
\narrowtext 
 

The Kondo model has been introduced to explain the transport and 
thermodynamic anomalies of dilute magnetic alloys~\cite{hewson} due to weak 
interaction between the conduction electrons and impurity spins. The model is 
defined by the Hamiltonian $H_K=J {\bf s}\cdot {\bf S}$, where ${\bf s}$ and 
${\bf S}$ are the spins of conduction and localized electrons, respectively,
and $J$ is the coupling constant. Our current understanding of Kondo systems 
is based on the solutions obtained by the peturbative 
scaling\cite{anderson.70}, numerical renormalization group\cite{wilson.75}, 
and Bethe Ansatz\cite{tsvelick.83}, which are usually derived for a 
featureless conduction band. The main result is the crossover from the 
high-temperature regime, where the impurity is magnetic and the conduction 
electrons are only weakly perturbed by the spins, to the strong coupling 
regime, where the magnetic moment is screened completely and the conduction 
states form a Fermi liquid~\cite{wilson.75}. The Kondo temperature separates 
these two regimes and is the only relevant energy scale. 

Kondo systems with a non-constant DOS or a high concentration of magnetic 
impurities show more complicated behavior. Assuming an electron-hole symmetry 
and a DOS with a power-law singularity at the Fermi level $E_F$, Whithoff and 
Fradkin~\cite{wf} found a critical coupling $J_{c}$ such that for $J>J_{c}$ 
the dilute Kondo systems scale to the strong coupling limit, while for 
$J<J_{c}$ the renormalized coupling decreases with temperature and the 
``usual'' Kondo screening does not occur. Here, we study the low-temperature 
Kondo effect in bad metals and  consider a lattice of magnetic ions coupled 
by exchange interaction to a conduction band with a minimum (or a gap or a 
pseudo-gap) close to $E_F$. We show that such a model generates more than one 
energy scale and explains the low-temperature behavior of Yb- and Eu-based 
intermetallic compounds, like  YbInCu$_4$~\cite{sarrao.99}, 
EuNi$_2$(Si$_{1-x}$Ge$_{x}$)$_2$~\cite{wada.97} or 
Eu(Pd$_{1-x}$Pt$_{x}$)$_2$Si$_2$~\cite{mitsuda.97}, which exhibit a 
valence-change transition. We also study the Kondo lattice with an enhanced 
DOS around $E_F$ and show that it explains the properties of 
YbAl$_3$~\cite{lawrence.02} in the coherent regime. 


The Kondo lattice Hamiltonian reads,
\begin{equation} 
H = \sum_{\langle i,j \rangle, \sigma}
(t_{ij}-\mu\delta_{ij})c_{i\sigma}^{\dagger}c_{j\sigma} 
+ 
\frac{J}{\cal N}\sum_{i}{\bf s}_{i}\cdot {\bf S}_{i}\;, 
\label{hamiltonian} 
\end{equation} 
where $t_{ij}$ is the hopping, $c_{i\sigma}^{\dagger}$ creates an electron of 
spin $\sigma=1,...,{\cal N}$ at site $i$, ${\bf s}_{i}$ is the c-electrons 
local spin density, $\mu$ is the chemical potential which determines the band 
filling, $n_c/2={\cal N}^{-1}\sum_\sigma\langle c_{i\sigma}^{\dagger}c_{i\sigma}\rangle$, and ${\bf S}_{i}$ are the local spins of SU(${\cal N}$) symmetry. 
We use the large-${\cal N}$ approach, write ${\bf S}_{i\sigma,\sigma'}=f^\dagger_{i\sigma} f_{i\sigma'}-\delta_{\sigma\sigma'}/2$, where 
$\sigma,\sigma'=1,...,{\cal N}$, and enforce the local constraints 
$\sum_{\sigma=1}^{\cal N} f^\dagger_{i\sigma} f_{i\sigma} = {\cal N}/2$ by 
the Lagrange-multipliers $i\lambda_i(\tau)$~\cite{largeN,BGG}. The quartic 
fermionic interactions in Eq.~(\ref{hamiltonian}) are decoupled by 
introducing Hubbard-Stratanovich transformations~\cite{BGG}, which lead in 
the limit $1/{\cal N}=0$ to the following set of equations for the 
hybridization parameter $r$ and the Lagrange multiplier $\lambda$,  
\begin{equation} 
\left\{ -\frac{r}{J}, \frac{1}{2}, \frac{n_c}{2} \right\} 
= -\left\{ G_{fc}, G_{f}, G_{c} 
\right\}(\tau=0^-)\;.  
\label{colkondo} 
\end{equation}  
Here, $G_{f}(\tau)=-\langle T f_0(\tau)f_0^{\dagger}(0)\rangle$ is the 
f-fermion Green's function, 
$G_{fc}(\tau)=-\langle T f_0(\tau)c_0^{\dagger}(0)\rangle$ is the mixed 
Green's function, and 
$G_{c}(\tau)=-\langle T c_0(\tau)c_0^{\dagger}(0)\rangle$ the c-electron 
Green's function, all of them being computed for an arbitrary site $i=0$ (In 
the paramagnetic state the spin index has been suppressed). Local Green's 
functions can be obtained from the free fermionic propagator 
${\cal G}_{f}(i\omega_n)=(i\omega_n+\lambda)^{-1}$, using the relations 
$G_{c}(i\omega_n)=G_{c}^{0}(i\omega_n+\mu-r^2{\cal G}_{f}(i\omega_n))$, 
$G_{f}(i\omega_n)={\cal G}_{f}(i\omega_n)+r^{2}{\cal G}_{f}^{2}(i\omega_n) 
G_{c}(i\omega_n)$ and $G_{fc}(i\omega_n)=r{\cal G}_{f}(i\omega_n)G_{c} 
(i\omega_n)$, in which $\omega_n$ are the Matsubara fermionic frequencies, 
and $G_{c}^{0}$ the non-interacting electronic Green 
function~\cite{largeN,BGG}. Typical band-shapes and fillings considered in 
this work are shown in Fig.~1.


The temperature $T_K$  at which the second order transition with the order 
parameter $r(T)$ takes place is obtained from Eqs.~(\ref{colkondo}) as a non 
trivial solution of the equation ($r=0$, $J>J_{c}$),
\begin{equation} 
\frac{2}{J} 
= 
\int_{-\infty}^{+\infty}d\omega 
\frac{{\rm tanh}\,[\omega/2T_K]}{\omega} 
\rho_0(\omega+\mu)\;.
\label{equation_TK} 
\end{equation} 
Here, $\rho_0(\omega)$ is the non-interacting DOS and the energy is measured 
with respect to the center of the conduction band, which is assumed to be 
symmetric and of half-width $D_0$. Assuming spin-rotation symmetry, we find 
that the large$-{\cal N}$ result for the lattice (\ref{equation_TK}) 
coincides with the solution of the scaling equations for the $SU(2)$ 
single-impurity Kondo model~\cite{wf}. The $T_K$ vanishes continuously at 
\begin{equation} 
\frac{2}{J_{c}} 
= 
\int_{-\infty}^{+\infty}d\omega 
\frac{\rho_0(\omega+\mu)}{|\omega |},   
\end{equation} 
and for $\rho_0(\mu)=0$ the model admits a finite $J_c$ and a quantum 
critical behavior. In the case of a DOS with a gap $\Delta_0\ll D_0$ around 
$E_F$ we find $J_{c}\propto D_0/ln[D_0/\Delta_0]$, while the pseudo-gap 
centered at $E_F$ and characterized by a single energy scale $D_0$ gives 
$J_{c}\propto D_0$. The power-law singularity 
$\rho_0(\omega)=A|\omega|^\gamma$ gives $J_{c}=2\gamma D_0/(\gamma+1)$, which 
is the Whithoff and Fradkin result~\cite{wf,vojta.02} (the constant $A$ 
follows from the normalization condition, $\int \rho_0(\omega)= 1$). However, 
for $\rho_0(\mu)\neq 0$ (see Figs.~1(a) and~1(b) or~1(e)), the integral 
diverges logarithmically and $J_{c}=0$.
\begin{figure}[t] 
\epsfxsize=3.3in 
\epsffile{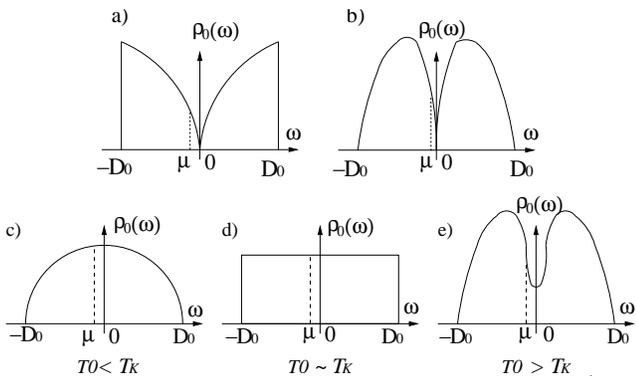} 
\label{fig} 
\caption{
Schematic non interacting DOS: 
singular (a and b), semi-elliptic (c), constant (d), and  
pseudo-gaped (e). 
As soon as $\rho_0(\omega=\mu)\neq 0$, the Kondo temperature $T_K$ is 
finite. Close to the half-filling, the shape of the DOS around 
$\mu$ is crucial for determining the Fermi-liquid temperature: 
(c) $\mu$ is close to a maximum of $\rho_0$ and $T^\star < T_K$. 
(e) $\mu$ is close to a minimum of $\rho_0$, and $T^\star > T_K$.  
(d) $\rho_0$ is constant and $T^\star\sim T_K$.  
} 
\end{figure}

Taking the non-constant DOS and solving ~Eq.(\ref{equation_TK}) in the weak 
coupling limit gives~\cite{BGG}, $T_K=F_K exp[-1/J\rho_0(\mu)]
$
where 
\begin{equation} 
F_K=\alpha\sqrt{D_{0}^{2}-\mu^2} 
exp\left( { 
\int_{-D_0-\mu}^{D_0-\mu} 
\frac{d\omega}{|\omega |} 
\frac{\rho_0(\mu+\omega)-\rho_0(\mu)}{2\rho_0(\mu)}}\right)\;, 
\label{expression_FK} 
\end{equation}
where $\alpha\approx 1.13$. This result is derived in ref.~(\cite{BGG}) for 
an even and analytic DOS but it also holds for $\rho_0(\omega)$ with an 
algebraic singularity close to $E_F$. The $T_K$ defines the Kondo temperature 
of the lattice; it characterizes the high-temperature behavior and gives the 
temperature at which the weak coupling behavior breaks-down due to the 
single-impurity singlet formation. The screening cloud at $T\approx T_K$ 
involves the electronic states in an energy interval $|\omega|\leq T_K$ 
around $E_F$ ($E_F$ is the non-interacting Fermi level of $n_c$ electrons 
having a ``small'' Fermi surface, FS). 

At low temperatures, Eqs.~(\ref{colkondo}) lead to a Fermi liquid (FL) ground 
state which is characterized by two energy scales, $T^{\star}$ and 
$T^{\star\star}$. The scale $T^{\star}$ is defined as the inverse of the 
quasi-particle DOS at the Fermi energy of a ``large FS'' which contains, at 
$T=0$, $n_c+n_f$ quasi-particles ($n_f$ is the occupancy of the localized 
$f$-orbitals, which is close to $1$ for $SU(2)$ Kondo systems). This scale is 
relevant for the $T\rightarrow 0$ properties and appears in the static spin 
susceptibility, $\chi_{loc}(T=0)\sim 1/T^{\star}$, or the specific heat 
coefficient $C/T\sim 1/T^{\star}$. The scale $T^{\star\star}$  characterizes 
the deformation  of the ``quasi-particle band'' with temperature; it defines 
the relevant energy scale for quasi-particle excitations and can be related 
to the temperature at which the coherence is lost and the quasiparticles 
Fermi surface destroyed. This  scale depends on all the electronic states 
between ``small'' and ``large'' FS, as well as on some hole states inside 
the ``small'' FS~\cite{BGG}. The system behaves as a FL provided 
$T/T_{FL}\ll 1$ where $T_{FL}$ is the smallest of the three temperatures 
defined above. One common situation arises for 
$T^{\star}\ll T^{\star\star}\ll T_K$, such that the band of quasiparticles 
is ``rigid'' and we have the FL laws, provided this rigid band does not have 
too much structure within the Fermi window. The other usual situation is 
$T^{\star\star}\ll T^{\star}\ll T_K$, such that the deviations from the FL 
laws are due to the temperature-induced deformation of the effective DOS, 
i.e. the break-down of the FL is due to the loss of coherence. A different 
behavior is expected in bad metals with only a few states around $E_F$ and 
$T_K\ll T^{\star\star}$ or $T_K\ll T^{\star}$. 

To study the relative magnitude of $T_K$, $T^{\star}$, and $T^{\star\star}$, 
we use the analytic expressions obtained from Eqs.~(\ref{colkondo}) and 
evaluate them close to half-filling~\cite{BGG,thesis}. This gives 
$T^{\star}\approx T^{\star\star}=F_0 exp[-1/J\rho_0(\mu)]$, and  
\begin{equation} 
F_0=(D_{0}+\mu ) 
exp\left( { 
\int_{-D_0-\mu}^{\epsilon_{F}^{>}-\mu} 
\frac{d\omega}{|\omega |} 
\frac{\rho_0(\mu+\omega)-\rho_0(\mu)}{\rho_0(\mu)} 
}\right)\;,  
\label{expression_F0} 
\end{equation} 
where $\epsilon_{F}^{>}$ is the  Fermi level corresponding to $(n_c+1)/2$ 
non-interacting fermions per spin component. The break-down of the FL laws is 
now due to quasi-particle excitations and a simultaneous destruction of the 
coherent state. Note, we assume that the system is not at half-filling exactly 
(where the ``large'' FS coincides with the Brillouin zone and we would get a 
Kondo insulator). But we also assume that the system is not too far from the 
particle-hole symmetry, because for small filling the 'exhaustion' gives rise 
to additional effects (see discussion by Nozi\`eres~\cite{Nozieres} 
and~\cite{noteexhaustion}). Both assumptions are fulfilled for $n_f\neq 1$ 
but not for $n_f=1$ or $n_f\simeq 0$. 
The expressions for $F_K$ and $F_0$ can be simplified to 
$F_K\approx \alpha D_0  
exp\left( { 
\int_{-D_0}^{D_0} 
\frac{d\omega}{|\omega |} 
\frac{\rho_0(\mu+\omega)-\rho_0(\mu)}{2\rho_0(\mu)} 
}\right)$  
and  
$F_0\approx D_0  
exp\left( { 
\int_{-D_0}^{D_0} 
\frac{d\omega}{|\omega |} 
\frac{\rho_0(\mu+\omega)-\rho_0(\mu)}{\rho_0(\mu)} 
}\right)$, such that $F_0\approx (F_K/\alpha)^2/D_0$. Thus, we have the 
relationships between the Kondo temperature and the FL scale
\begin{equation} 
T^\star=\frac{F_K}{\alpha^2 D_0}T_K\;, 
\end{equation} 
regardless of the specific form  of $\rho_0$. A constant $\rho_0$ gives 
$F_K/D_0=1$ and $T^\star\sim T_K$, which explains the $T/T_K$ scaling 
observed in many heavy fermion compounds.

We consider now in more detail the effects due to the shape-variation of 
$\rho_0(\omega)$. If $\mu$ is close to a local maximum of $\rho_0(\omega)$ 
the integrand in $F_K$ is mainly negative, such that $F_K/D_0<1$ and 
$T^{\star}\leq T_K$, as found in systems showing 
`protracted screening'~\cite{jarrell,lawrence.02} (exhaustion effects can 
also reduce $T^{\star}/T_K$~\cite{Nozieres}and see~\cite{noteexhaustion}). 
However, if $\mu$ is close to a local minimum (see Fig.~1(e)) one finds 
$T_K \ll T^\star$. Note, the relative magnitude of $T^\star$ and $T_K$ does 
not depend on $T_K$ (or $J$) but is related to the convexity of 
$\rho_0(\omega)$ around $\omega=\mu$. An intuitive interpretation is provided 
by the following argument. The incoherent Kondo cloud forms at $T\approx T_K$ 
and involves only a few states around $E_F$. But the energy scale $T^{\star}$ 
depends on the density of quasiparticle states at the Fermi level of the 
``large'' FS, and the energy $T^{\star\star}$ involves all the states between 
the ``small'' and the ``large'' FS, and some additional holes inside the 
``small'' FS~\cite{BGG,thesis}. Thus, for $\mu$ close to the minimum of 
$\rho_0(\omega)$, the integrals for $T^{\star}$ and $T^{\star\star}$ involve
the DOS which is much larger than the one used to evaluate $T_K$, and 
$T_K\ll T^\star$ follows. 


Next, we study the effects of an external magnetic field $H$ applied to the 
spin system at $T=0$. Within the large-${\cal N}$ approach, we obtain a 
transition between the Kondo phase ($r\neq 0$, ``large FS'') and the 
ferromagnetic spin lattice decoupled from the paramagnetic metal ($r=0$, 
``small'' FS). The critical field $H_{c}$ is given by the relation 
\begin{equation} 
\frac{2}{J} 
= 
\int_{-\infty}^{+\infty}d\omega 
{\rm tanh}\left[ \frac{\omega}{2T}\right] 
\frac{\omega }{\omega^2-H_{c}^2} 
\rho_0(\omega+\mu), 
\label{equation_TKetH} 
\end{equation} 
which shows that the logarithmic divergence of~Eq.(\ref{equation_TK}) 
disappears, and $J_c\neq 0$, as soon as $H_{c}\neq 0$. In the weak coupling 
limit, $J\ll D_0$, we find $H_{c0}=\alpha^{-1}F_Kexp[-1/J\rho_0(\mu)]$ and 
obtain, at $T=0$, the universal relation $H_{c0}= k_B T_K/\mu_B \alpha$.
Thus, in Kondo lattices with a non-constant DOS we can expect two different 
types of the zero-temperature magnetization $m(H)$. For $T^\star\ll T_K$ and 
$ H \ll H_{c}$, the magnetization rises initially as $m(H)\propto H/T^\star$ 
but for higher fields the slope is reduced and eventually, around 
$H^\star\propto k_B T^\star/\mu_B \ll k_B T_K/\mu_B $, the $m(H)$ saturates. 
For $H^\star \ll H\ll H_c$ we are dealing with a saturated FL ($r\neq 0$). In 
the opposite case, $T_K\ll T^\star$ and $H_{c}\ll H^\star$, the low field 
limit still gives $m(H)\propto H/T^\star$. But for higher fields the slope of 
$m(H)$ rises rapidly and at about $H_{c}\propto k_B T_K/\mu_B$ there is a 
meta-magnetic transition into the spin-disordered (r=0) phase. Of course, 
this simple considerations should be corrected for direct and indirect 
effects due to the conducting sea. 

Solving Eq.~(\ref{equation_TKetH}) for a constant DOS we find the critical 
line $[H_c(T)/H_{c0}]^2+[T/T_K]^2=1$, which holds for any $J$ and $n_c$. The 
same relation is also found at half-filling, for any DOS. In general, we 
expect a `nearly' universal phase boundary, with some small deviations due to 
the structure of the  DOS {\it and} the particle-hole asymmetry. 

 
We can use these results to discuss the anomalous behavior of the Yb- and 
Eu-based intermetallics that we have mentioned before. In YbInCu$_4$, which 
is a typical example\cite{sarrao.99}, the valence-change transition takes 
place, at ambient pressure, at $T_v\approx 40K$. Above $T_v$, one finds the 
Yb ions in a 3$^+$ state, behaving as 'almost free spins', and the transport 
properties of a bad metal. That is, the linear and non-linear magnetic 
response of the paramagnetic phase are very well explained by the crystal 
field theory of independent f-states, while the electrical resistance is very 
high, there are no logarithmic terms, and the fields of up to 40 Tesla do not 
produce any significant effects. The Kondo scale determined from the 
high-temperature properties seems to be very small. Below $T_v$, the valence 
state of Yb ions changes to 2.9$^+$ and the system behaves as a FL with a 
high characteristic scale, $T^\star\approx 500K$~\cite{silverprlsarrao}, such 
that the physical quantities are nearly temperature-independent. The optical 
conductivity\cite{garner_2000}, Hall effect\cite{figuroa_1998} and the 
thermoelectric power\cite{ocko_2002} indicate a major reconstruction of the 
conduction band: a bad metal with chemical potential close to the pseudo-gap 
(or small gap) transforms at $T_v$ into a good metal with a large FL scale. 
The FL ground state can be destroyed by a magnetic field of about 
$H_{c0}\leq 40$ Tesla, which induces a metamagnetic transition and restores a 
bad metal. The valence-change transition in Eu-based intermetallics is 
similar~\cite{wada.97,mitsuda.97}, except the transition temperatures and the 
critical fields are higher ($T_v\geq 100$ K and $H_c\geq 50$ Tesla), and the 
Eu ions undergo an almost complete valence change  from 2+ ($f^7$) to 3+ 
($f^6$) state. 

We explain such a behavior of YbInCu$_4$-like intermetallics by the proximity 
of $\mu$ to the pseudo-gap, which reduces $T_K$ and gives 
$T^\star\gg T_K\propto H_c$. If we take $T_v \propto T_K$, the Kondo lattice 
model describes the valence-change as a transition from a bad metal with a 
``small FS'' to a good metal with hybridized f-states and ``large FS''. Above 
$T_v$, the Yb ions form a disordered spin-lattice which is decoupled from the 
conduction band, while below $T_v$ the f-electrons participate in the 
``heavy'' quasiparticle band with a ``large'' FS. Since $T^\star$ is not 
affected by the pseudo-gap, and $T^\star\gg T_v$, all the properties of the 
system below $T_v$ are are nearly temperature-independent. If we estimate the 
FL corrections to the $T=0$ value of the magnetic susceptibility or the 
electric resistivity up to ${\cal O}[(T/T^\star)^2]$ and assume 
$T^\star/T_K\leq 10$ (as indicated by the data), the maximum relative 
deviation at $T\sim T_v$ is  $1$\%. The large$-{\cal N}$ solution explains 
the metamagnetic transition at the critical field, 
$H_{c}\propto T_v\propto T_K$, and gives the universal ratio 
$H_{c}/T_v={\cal O}(1)$  but can not describe the properties of 
YbInCu$_4$-like systems above $T_v$. The proper description of the 
high-temperature phase might require a Coulomb repulsion between the f-states 
and the conduction band, i.e. the free electron states of the 
high-temperature phase should be replaced by the over-damped states of a 
small-gap Mott-Hubbard insulator. A gap or a pseudo-gap is also required to 
explain the absence of the Kondo effect in the high-temperatutre phase, 
despite the presence of a local moment at each lattice site~\cite{veljko}. 

The YbAl$_3$ anomalies are of a different type but also exhibit more than one 
low-temperature energy scale. The data~\cite{lawrence.02} show that for 
$T\geq T^\star\simeq 40$ K the coherence is lost but the strong coupling 
features persist up to the highest temperatures measured, i.e.,  the magnetic 
moment remains quenched ($T_K\geq 650$ K). The de Haas - van Alphen 
experiments in the coherent regime show~\cite{lawrence.02} that the 
high-field dHvA mass is greatly reduced with respect to the low-field mass, 
and that the critical field is about $H^\star \simeq k_B T^\star/\mu_B$. 
However, there is hardly any change in the Fermi surface for 
$H\leq H^{\star}$. Using the Kondo lattice model close to half-filling and 
assuming that the chemical potential is close to the maximum of the DOS, we 
find $T^\star\ll T_K$ and $H^\star \ll k_B T^\star/\mu_B$. Thus, we have 
$m(H)\simeq H/T^\star$ for low fields and $m(H)\propto$ const for 
$H\geq T^\star$, i.e. the saturation magnetization is approached in the usual 
Fermi liquid fashion. Since $H^\star\simeq T^\star \ll T_K$, the order 
parameter is finite and the system can be viewed as a polarized heavy FL with 
a ``large FS''. 

In summary, we studied the low-temperature properties of the Kondo lattice 
model with a non-constant DOS, using large-${\cal N}$ formalism. Close to the 
particle-hole symmetry, the ground state is a Fermi-liquid, except for 
$\rho_0(\mu)=0$, in which case $T_K=0$ for $J\leq J_c$, where $J_c$ is a 
quantum critical point. This generalizes the work of Withoff and 
Fradkin~\cite{wf} which found the quantum critical behavior for a Kondo 
impurity model with a singular DOS. We find that $T_K$ is finite but reduced 
for a conduction band with only a few states around to $\mu$. Since the FL 
parameter $T^\star$ of the Kondo lattice is much less affected by the 
pseudo-gap than $T_K$, the values of $T_K$ and  $T^\star$ can differ by more 
than one order of magnitude. If that is the case, the physical properties of 
the FL phase are nearly constant. We also find that a DOS with a peak close 
to $\mu$ would reverse the relative magnitude of $T_K$ and  $T^\star$. 
Considering the magnetic field effects, we find the critical value 
$H_{c}\simeq (k_B/\mu_B) T_K$ which separates the Kondo phase (Fermi liquid 
with a large Fermi surface and hybridized bands) from the Kondo-free phase 
(bad metal with a pseudo-gap and a ``small FS'', decoupled from the f-spins). 
The low-field response of the Kondo lattice is linear and proportional to 
$1/T^\star$, while the high field behavior sets in at about $H_c\propto T_K$ 
or $H^\star\propto T^\star$, whichever is smaller. However, the non-linear 
corrections go in the opposite directions for $T^\star \gg T_K$  and 
$T^\star \ll T_K$ systems. In the former case there is a metamagnetic 
transition at $H_c$, while in the latter one finds at $H^\star$ the usual 
paramagnetic saturation. These Kondo lattice results explain different 
behavior and various energy scales observed in a valence fluctuator such as 
YbAl$_3$, and in systems with a valence-change transition, such as 
YbInCu$_4$, EuNi$_2$(Si$_{1-x}$Ge$_{x}$)$_2$ or 
Eu(Pd$_{1-x}$Pt$_{x}$)$_2$Si$_2$.

We thank C. Castellani, C. Di Castro, D.R. Grempel, M. Grilli, E. Kats and 
P. Nozi\`eres for useful discussions. V.Z. thanks the hospitality of the 
Institute Laue-Langevin, where the majority of this work has been completed, 
and acknowledges the support from the Swiss National Science Foundation, 
Grant No. 7KRPJ65554. S.B. acknowledges financial support from the FERLIN 
program of the European Science Foundation.

\vspace {-0.5 truecm} 


\begin{references}
\bibitem{noteexhaustion} 
The exhaustion problem discussed by 
Nozi\`eres~\cite{Nozieres} and latter in Ref.~\cite{BGG} within the 
large$-{\cal N}$ formalism, concerns the limit  
$n_c\ll 1$ (i.e. $\mu\rightarrow -D_0$),  
for which one gets $T^*\ll T_K$ due to the pre-factors  
$\sqrt{D_{0}^{2}-\mu^2}$ and $(D_{0}+\mu)$ occurring in  
Eq.~(\ref{expression_FK}) and Eq.~(\ref{expression_F0}) respectively  
(see also Fig.~1 for $\mu$ close to the bottom 
of the band).
\end{references}
\end{document}